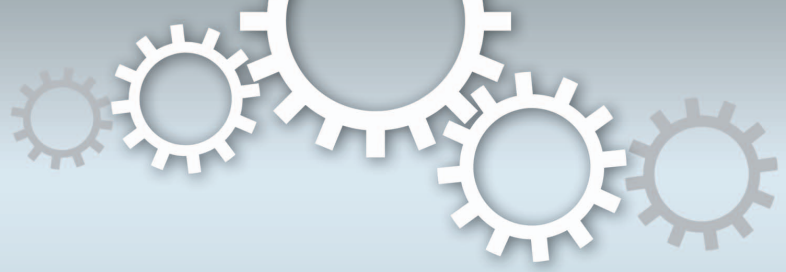

# SCIENTIFIC REPORTS

# Reversible Modulation of Spontaneous Emission by Strain in Silicon Nanowires

Daryoush Shiri[1], Amit Verma[2], C. R. Selvakumar[1] & M. P. Anantram[3]

[1]Department of Electrical and Computer Engineering, University of Waterloo, Waterloo, Ontario N2L 3G1, Canada, [2]Department of Electrical Engineering and Computer Science, Texas A&M University – Kingsville, Kingsville, Texas 78363, USA, [3]Department of Electrical Engineering, University of Washington, Seattle, Washington 98105-2500, USA and Department of Physics, University of Waterloo, Waterloo, Ontario N2L 3G1, Canada.

We computationally study the effect of uniaxial strain in modulating the spontaneous emission of photons in silicon nanowires. Our main finding is that a one to two orders of magnitude change in spontaneous emission time occurs due to two distinct mechanisms: (A) Change in wave function symmetry, where within the direct bandgap regime, strain changes the symmetry of wave functions, which in turn leads to a large change of optical dipole matrix element. (B) Direct to indirect bandgap transition which makes the spontaneous photon emission to be of a slow second order process mediated by phonons. This feature uniquely occurs in silicon nanowires while in bulk silicon there is no change of optical properties under any reasonable amount of strain. These results promise new applications of silicon nanowires as optoelectronic devices including a mechanism for lasing. Our results are verifiable using existing experimental techniques of applying strain to nanowires.

P otential advantages of silicon nanowires (SiNWs), such as quantum confinement, large surface-to-volume ratio, adjustable bandgap, sensitivity of electronic properties to surface ligands and mechanical excitation and compatibility with mainstream silicon technology have resulted in a flurry of experimental and theoretical investigations of these nano-structures. Over the years SiNWs have been explored for use in transistors[1,2], logic circuits[3] and memory[4], quantum computing[5], chemical[6] and biological[7] sensors, piezo-resistive sensor[8], nano mechanical resonator[9] and thermoelectric converters[10,11]. We are also witnessing the utilization of SiNWs in optoelectronic applications, such as in solar cells[12,13], photodetectors[14,15], and avalanche photodiodes[16,17]. The basic properties and applications of Si nanowires are discussed further in references[18,19]. Efficient light emission in SiNWs requires direct bandgap and symmetry allowed optical transition between conduction and valence states. In other word to have a nonzero optical transition matrix element which is defined as

$$\langle \psi_c | \boldsymbol{r} | \psi_v \rangle = \int \psi_c^*(\boldsymbol{r}) . \boldsymbol{r} . \psi_v(\boldsymbol{r}) d\boldsymbol{r} \qquad (1)$$

the integrand should have an even symmetry. $\Psi_c$, $\Psi_v$ and $\boldsymbol{r}$ represent conduction band state (wave function), valence band state and position operator, respectively. In bulk silicon and large diameter SiNWs the conduction band minimum and valence band maximum have different values of momentum within the Brillouin Zone (BZ) of the crystal. Spontaneous emission of a photon which arises from electron-hole recombination is a momentum conserving process. As photons cannot provide the momentum difference in these materials, a phonon absorption/emission is necessary, making it a weaker second order process. Vital to realizing SiNW-based light emitting devices, narrow diameter [110] and [100] SiNWs are direct bandgap. This arises from folding of four degenerate indirect X conduction valleys of bulk silicon into the BZ center (Γ point) due to confinement in transverse directions[20]. In addition to direct bandgap, the possibility of adjusting the bandgap with mechanical strain provides a new degree of freedom for SiNWs. Computational studies using both tight binding[21–23] and Density Functional Theory (DFT)[24–26] have shown that axial strain changes the bandgap of narrow SiNWs. Additionally it causes direct to indirect bandgap conversion.

On the experimental side, recent research points to promising directions for applications involving strain in nano-structures. For example, strain that is generated by an acoustic wave can modulate the energy levels of an artificial atom (quantum dot) and initiate lasing inside a Fabry-Perot device[27]. Similarly, the energy levels of a phosphorous atom embedded in a SiGe super lattice can be modulated by acoustic waves travelling back and forth in the super lattice[28]. There is evidence of modulation of threshold voltage in a transistor by transverse as well as longitudinal strains applied to the SiNW-based channel[29]. Carrier mobility enhancement due to residual tensile strain from the oxide layer in Gate-All-Around (GAA) 5 nm thick SiNWs have been reported[2]. He and Yang[8] have reported a large diameter ($d$)SiNW ($d>50$ nm) bridge in which the piezo-resistivity is varied by deformation of the substrate. Deforming an elastomeric substrate can apply $\pm 3\%$ strain to the buckled SiNWs grown on the substrate[30]. Strain induced by the cladding of SiNWs causes a blue shift in the UV Photoluminescence (PL) spectrum[31]. Similarly the observed red shift in the PL spectrum of 2–9 nm thick SiNWs is also attributed to the radial strain induced by oxide cladding[32].

Motivated by the aforementioned experiments, the scope of our work is to address this question: Is it possible to modulate the spontaneous emission time of electrons in SiNWs using axial strain? We demonstrate for the first time that applying strain can change the spontaneous emission time by more than one order of magnitude. The underlying reasons are a change in the symmetry of electronic wave functions and a change in the nature of the bandgap (direct or indirect), with strain. Qualitatively we might expect that converting direct bandgap to indirect bandgap hinders light emission from a SiNW because the emission of a photon is now a slow second order phonon-mediated process. We use Density Functional Theory (DFT) and tight binding methods to quantify the change of spontaneous emission time (see **Methods** section).

## Results

The cross section and electronic structure of an unstrained 1.7 nm [110] SiNW are shown in Fig. 1(a,b). All SiNWs in this work have direct bandgap at 0% strain (Fig. 1b), where the bandgap is observed to be inversely proportional to the SiNW diameter ($d$). For example the bandgap of 1.26 nm, 1.7 nm, 2.3 nm and 3.1 nm diameter SiNWs are 2.24 eV, 1.74 eV, 1.58 eV and 1.477 eV, respectively. The bandgap values are calculated using the tight binding framework after the atomic structure of the nanowires is relaxed by DFT. This effect of SiNW diameter on the bandgap agrees with experimental results using scanning tunneling spectroscopy (STS)[33] and the observed blue shift in PL spectrum of small SiNWs[31]. It is also observed that by increasing the diameter of SiNWs the difference between direct and indirect conduction band minima (band offset or $\Delta E_{cmin}$ in Fig. 1b) decreases. This aspect is important for this work, since it implies a corresponding decrease in the value of compressive strain (threshold strain) required to change the bandgap from direct to indirect[21,22]. The values of the threshold strain for 1.7 nm, 2.3 nm and 3.1 nm [110] SiNWs are $-5\%$, $-4\%$ and $-3\%$, respectively.

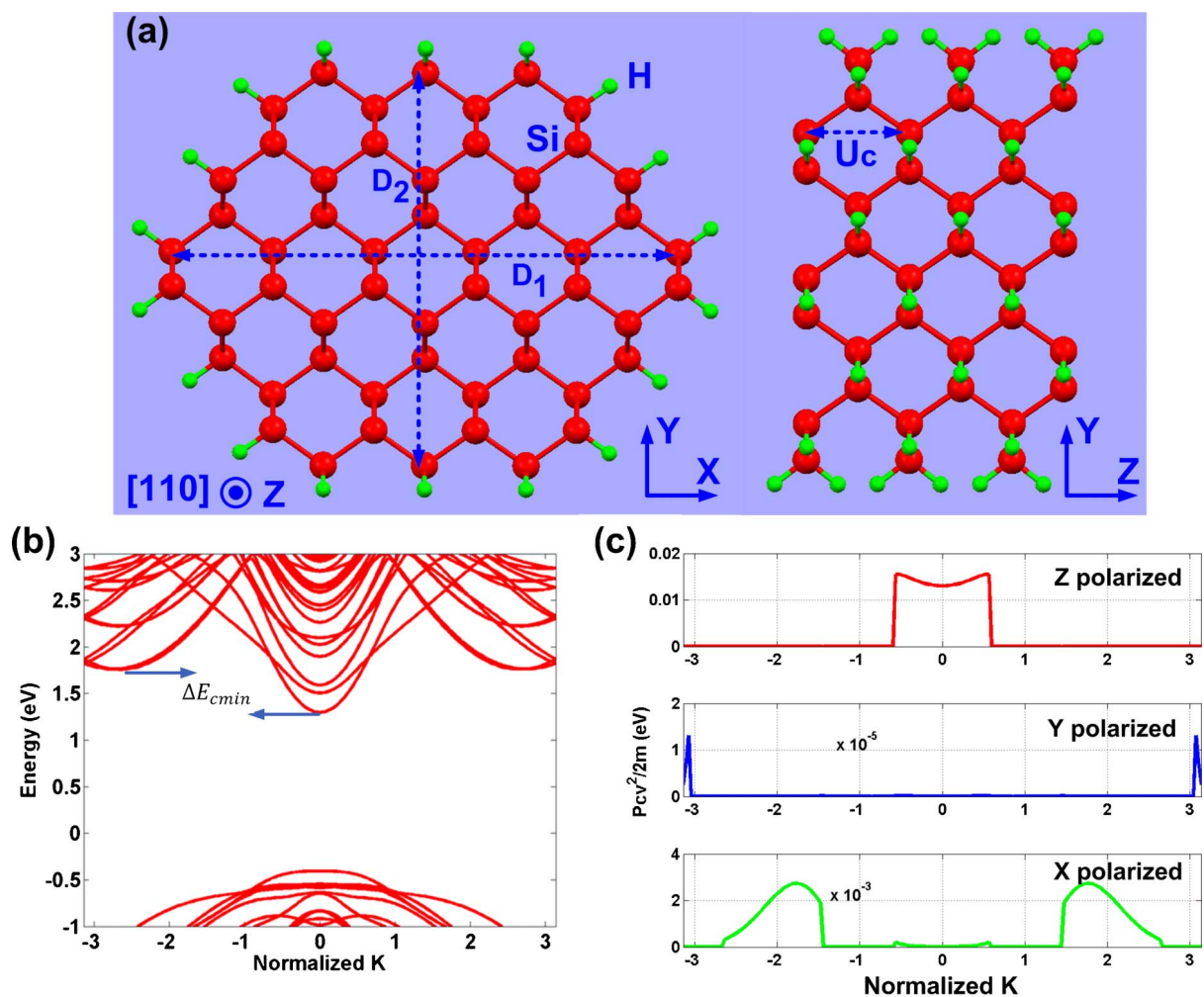

**Figure 1 | Cross section, electronic structure and momentum matrix element.** (a: left) Cross section of a [110] SiNW with the diameter (d) of 1.7 nm in xy plane. Diameter ($d$) is defined as average of large (D1) and small (D2) diameters. Right panel shows the side view of two unit cells along the length of nanowire (z) with unit cell length of $U_c$. Dark and bright atoms represent Si and H atoms, respectively. (b) Electronic structure of the 1.7 nm diameter [110] SiNW, which shows direct bandgap i.e. conduction/valence band minimum/maximum reside on the BZ center (K=0). Band offset, $\Delta E_{cmin}$, is smaller for larger diameter nanowires. (c) Normalized momentum matrix element, $P^2_{cv}/2m$ (in eV) between conduction and valence band along BZ. Red, Blue and green correspond to Z, Y and X polarizations of emitted photon.

**Direct bandgap regime.** If the conduction and valence states in the center of the 1st BZ are called initial (*i*) and final (*f*) states, respectively, the spontaneous emission rate or $1/\tau_{spon}$ can be written as

$$\frac{1}{\tau_{spon}} = \frac{2e^2 n_r}{3m^2\varepsilon_0 c^3 \hbar}\omega_{if}\left|\left\langle\psi_i\right|\hat{\mathbf{e}}\cdot\mathbf{P}\left|\psi_f\right\rangle\right|^2 \quad (2)$$

where $m$, $e$, $\varepsilon_o$, $c$, $\hbar$ are free mass of electron, magnitude of electronic charge, free space dielectric permittivity, speed of light in vacuum, and Planck's constant, respectively. $n_r$ is the refractive index of SiNW, which is assumed to have the same value as of bulk silicon ($n_r$=3.44) within the optical spectrum of interest. For all nanowires in this work, the bandgap lies in the 1–2 eV range, and the maximum change of bandgap due to strain can be as large as 500 meV (in a strain window of ±5%). Therefore assuming a constant value for the refractive index causes up to a 14% change in our calculated results (which is a small effect as we are discussing more than an order of magnitude change in $\tau_{spon}$). The quantity of $\omega_{if}$ is the frequency of the emitted photon, which is defined as $\Delta E_{if}/\hbar$. Here $\Delta E_{if}$ is the energy difference between initial (*i*) and final (*f*) states at the BZ center, which is the bandgap $E_g$. $P^2_{if} = |<\Psi_i|\hat{e}.P|\Psi_f>|^2$ is the matrix element of the momentum operator **P**, between initial (conduction) and final (valence) states. $\Psi_i$ and $\Psi_f$ are the wave functions of the initial (conduction) and final (valence) states. $\hat{e}$ is the direction of polarization. The value of momentum matrix element normalized to electronic mass, $P^2_{cv}/2m$, between conduction ($\Psi_i$) and valence ($\Psi_f$) bands along the BZ is plotted for an unstrained 1.7 nm diameter [110] SiNW (Fig. 1c). Corresponding to each direction of photon polarization (x, y and z), there are 3 different values for $P^2_{cv}$, which in turn leads to three different values for $\tau_{spon}$ according to equation (2). As Fig. 1c suggests, the momentum matrix element corresponding to z-polarized photons is significantly larger than the corresponding amounts for x- and y-polarized cases at BZ center. This difference manifests itself as a large spontaneous emission rate for z-polarized photons. This implies that the emitted light from the SiNW is mostly polarized along the length of the nanowire. In other words, if the average rate of spontaneous emission is defined as $\tau^{-1}_{avg} = \tau^{-1}_x + \tau^{-1}_y + \tau^{-1}_z$, then the degree of anisotropy or $\tau_{avg}/\tau_z$ has a value close to unity. Our approach of neglecting dielectric mismatch is justified because local field effects do not cause a significant change in the dielectric function of nanowire for z-polarized light[34,35].

The squared momentum matrix element in equation (2) is inversely proportional to the area of the nanowire or $d^2$, which can be explained using the particle in a box model[36,37]. Combination of this effect and the bandgap change with diameter results in the direct proportionality of spontaneous emission time to the cross sectional area. For 1.7 nm, 2.3 nm and 3.1 nm diameter [110] SiNWs, the value of $\tau_{avg}$ is calculated to be $1.47\times10^{-7}$ s, $2.3\times10^{-7}$ s, and $4\times10^{-7}$ s, respectively. For the SiNWs investigated in this work, the bandgap remains direct for most of the strain values in the ±5% range. For 1.7 nm, 2.3 nm and 3.1 nm [110] SiNWs the bandgap becomes indirect once the compressive strain reaches −5%, −4% and −3%, respectively[21,22].

As long as the bandgap is direct, photon emission is a first order process and its rate is governed by equation (2). The average spontaneous emission times for [110] axially aligned SiNWs in this study are tabulated in Table 1. As can be seen in Table 1, compressive strain leads to an increase of spontaneous emission time by one to two orders of magnitude. This is due to the movement of sub bands in the compressive strain regime. As pictured in the graphics of Fig. 2a, the rise of the second valence sub band (V2) due to its anti-bonding nature is more dominant than the rate with which the first valence sub band (V1) rises or the conduction sub band (C2) falls. As a result, V2 determines the new highest valence band. The aforementioned mechanism can be further understood by looking at Fig. 2b that shows the normalized probability density ($|\Psi|^2$) of conduction and valence states at BZ center. Comparing the valence/conduction bands (VB/CB) at 0% and −2% strain values shows that the dominant change is due to the valence band symmetry change induced by compressive strain (e.g. −2%). Left panel of Fig. 2b shows that the newly raised valence sub band (V2) has different wave function symmetry as opposed to the centro-symmetric nature of valence band V1 at 0% and +2% strains. Therefore the matrix element, $<\Psi_c|r|\Psi_v>$, changes accordingly and modulates the spontaneous emission time (rate) through equation (2). Comparing wave function symmetries of valence and conduction bands at strain values of 0% and +2%, as in Fig. 2b (center and right panel), further illustrates why the spontaneous emission time is almost unchanged within this tensile strain regime.

Table 1 | Average spontaneous emission time, $\tau_{avg}$ (s) vs. diameter (nm) of [110] SiNWs. Although all nanowires in these strain values have direct bandgap, the change of $\tau_{avg}$ with compressive strain is mainly due to valence sub band exchange

| Average spontaneous emission time, $\tau_{avg}$ (s) | | | |
|---|---|---|---|
| Strain (%) | 1.7 nm | 2.3 nm | 3.1 nm |
| +2% | $1.49\times10^{-7}$ s | $2.50\times10^{-7}$ s | $3.77\times10^{-7}$ s |
| 0% | $1.47\times10^{-7}$ s | $2.30\times10^{-7}$ s | $4.00\times10^{-7}$ s |
| −2% | $2.32\times10^{-6}$ s | $3.12\times10^{-6}$ s | $6.32\times10^{-5}$ s |

**Direct to indirect bandgap conversion.** The second kind of strain induced change of spontaneous emission arises from a direct to an indirect bandgap conversion. Fig. 3a elucidates this mechanism, in which, a compressive strain lowers the indirect conduction sub band C2, resulting in an indirect bandgap. Fig. 3 (b,c) shows the band structure of a 3.1 nm SiNW at 0% and −5% strains, respectively. As can be seen in Fig. 3c the heavy (H) or high effective mass sub-band (C2) has a lower energy than the light (L) or low effective mass direct conduction sub band (C1). The energy difference between two conduction band minima or band offset ($\Delta\Omega$) [Fig. 3a (right panel)] matters in determining the order of phonon-mediated spontaneous emission process. As mentioned before, $\Delta E_{cmin}$ (Fig. 1b) is inversely proportional to the nanowire diameter. As a result, a large diameter SiNW has a larger value of $\Delta\Omega$. For 1.7 nm, 2.3 nm and 3.1 nm diameter [110] SiNWs, $\Delta\Omega$ is 21 meV, 56 meV and 80 meV, respectively at −5% strain. When $\Delta\Omega$ is less than the Debye energy of LA phonons ($E_{Debye}$ = 54 meV) many secondary states in direct conduction sub band are available to which an electron can scatter from indirect sub band by absorbing a LA phonon. Alternatively, when $\Delta\Omega$ is less than the maximum energy of LO phonons, $E_{LO}$ = 63 meV, a few secondary states in direct conduction sub band can be found to which an electron can scatter from indirect sub band by absorbing a LO phonon. This implies that if the secondary state within the direct conduction sub band C1 is not at the BZ center, the only possible 1st order transition is due to LA phonon absorption. Otherwise both LA and LO phonon absorption processes will contribute in the 1st order inter-sub band scattering event. The process of finding secondary states can be understood by recalling that LO and LA phonons are modelled as dispersion-less and a linear dispersion around BZ center, respectively[38].

**First order transition.** Within the regime of small $\Delta\Omega$, the spontaneous emission is modeled as two consecutive 1st order processes as shown in Fig. 4a (top panel), the first of which is an electron-phonon scattering event from indirect sub band minimum to the direct sub band minimum (via absorption of a phonon), while the second is a direct transition from conduction band to valence sub band maximum via emission of a photon. With this model, the total spontaneous emission rate is limited by the slower of the two processes, which is the optical transition.

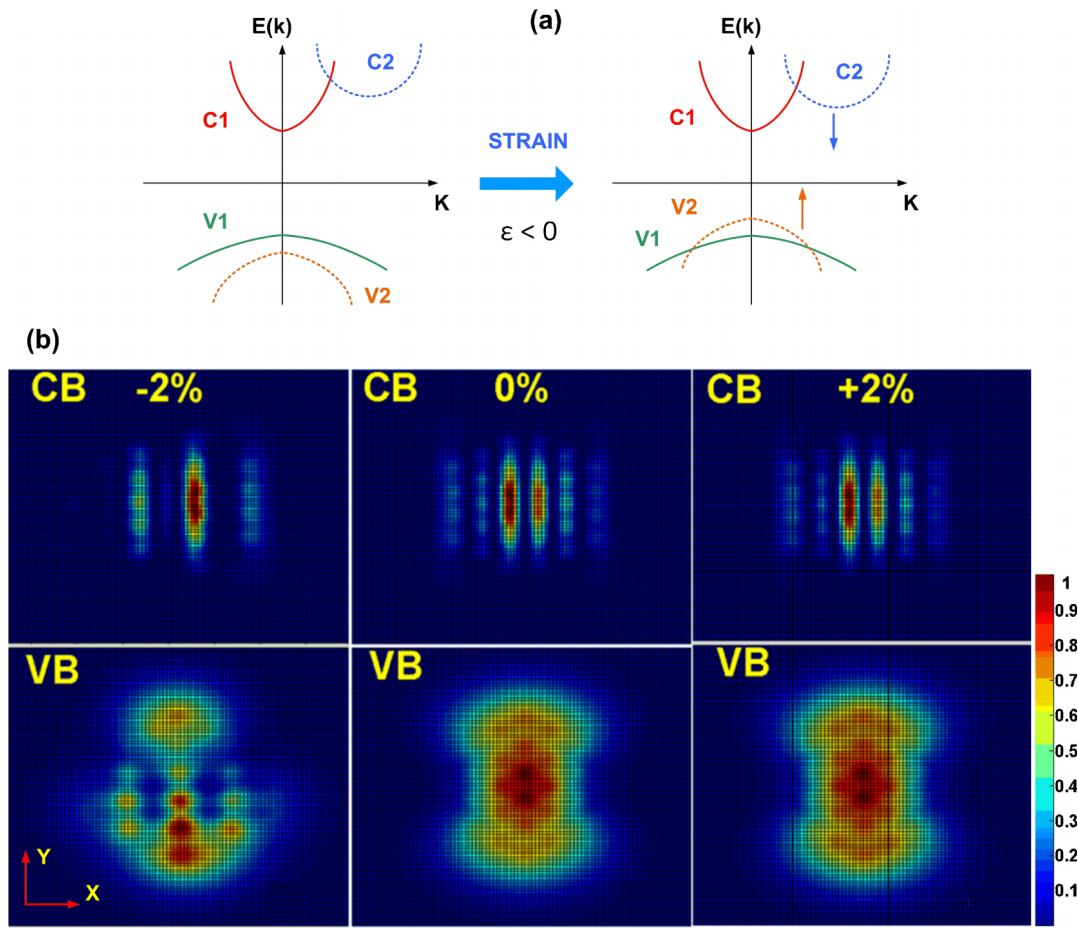

**Figure 2 | Wave function symmetry change with strain.** (a) Effect of compressive strain on second valence sub band (V2) which results in the change of wave function symmetry. From left to right, it can be seen that compressive strain raises the energy of V2 faster than it lowers the energy of C2. (b) Normalized squared value of wave function ($|\Psi|^2$) in the cross sectional plane of a 1.7 nm [110] SiNW. From left to right strain values are −2%, 0% and +2%, respectively. Valence and conduction band states (VB and CB) are at BZ center. As can be seen in the left panel (for −2% strain) the change of symmetry is more pronounced for valence sub band.

Fig. 4b shows the inter-sub band electron-phonon scattering rates for electrons making transition between the indirect sub band minimum at $K_z = 0.8\pi/U_c$ to the direct sub band minimum at $K_z = 0$, where $U_c$ is the nanowire unit cell length. One can observe that the scattering rate due to LO phonons is 2 orders of magnitude higher than the scattering due to LA phonons. Therefore when the nanowire bandgap is indirect with $\Delta\Omega < E_{Debye} = 54$ meV or $\Delta\Omega < E_{LO} = 63$ meV, $\tau_{spon}$ is determined by the slower (optical transition) process. Recalling Fig. 4a (top panel) and the rates given in Fig. 4b, $\tau_{spon}$ is calculated by equation (2). This results in spontaneous emission times that are comparable with that of direct bandgap nanowires i.e. in the $10^{-5}$–$10^{-7}$ second range, depending on the value of $P^2_{cv}$.

**Second order transition.** For indirect bandgap SiNWs in which $\Delta\Omega > E_{Debye}/E_{LO}$ (e.g. 3.1 nm diameter [110] SiNW at −5% strain), the spontaneous emission is possible through two second order processes that correspond to LA and LO phonons [Fig. 4a (bottom panel)]. The rate of both processes is given by second order perturbation theory. Equation (3) shows the spontaneous emission lifetime of an electron in indirect conduction sub band. The recombination of this electron with a hole residing at BZ center is possible through virtual transitions to intermediate states $\Psi_m$ in the first conduction band (via phonon emission and absorption)

$$\frac{1}{\tau_{spon}} = \frac{2\pi}{\hbar} \sum_{K_f} \sum_{\mathbf{k'},\sigma} \sum_{\mathbf{q},\lambda} \left| \sum_{K_m} \frac{\langle \psi_f | H_{e-R} | \psi_m \rangle \langle \psi_m | H_{e-P} | \psi_i \rangle}{E_i - E_m} \right| \delta(E_i - E_f) \delta_{k_i - k_m, \pm \mathbf{q}} F(E_f) \quad (3)$$

Here, $H_{e-R}$ and $H_{e-P}$ are electron-photon (radiation) and electron-phonon interaction Hamiltonians, respectively. The innermost summation is over all intermediate states. Indices *i*, *m* and *f* correspond to initial (indirect conduction band minimum), intermediate (within the 1st conduction band) and final (within the last valence band) states. $\Psi_i$, $\Psi_m$ and $\Psi_f$ are Bloch states of initial (*i*), intermediate (*m*) and final (*f*) states, respectively. As shown in Fig. 4c, this transition is possible through two processes of A→B→C and A→D→C. The process of A→D→C is not included in equation (3) since for this transition, the denominator of the inner most terms i.e. $E_i$-$E_m$ is large (>3 eV). Hence the contribution of A→D→C processes is negligible. Two outer summations are performed over phonon wave vector and branches (**q**, λ) and photon wave-vector and polarizations (**k'**, σ), respectively. The outermost summation is over all final states within the valence band, which are weighted by Fermi-Dirac occupancy written as $F(E_f)$. Dirac and Krönecker delta functions impose energy and momentum conservation, respectively. After converting summations to integration and calculating the interaction Hamiltonian matrix elements[39,40], the following equation results for spontaneous emission time including LA phonon

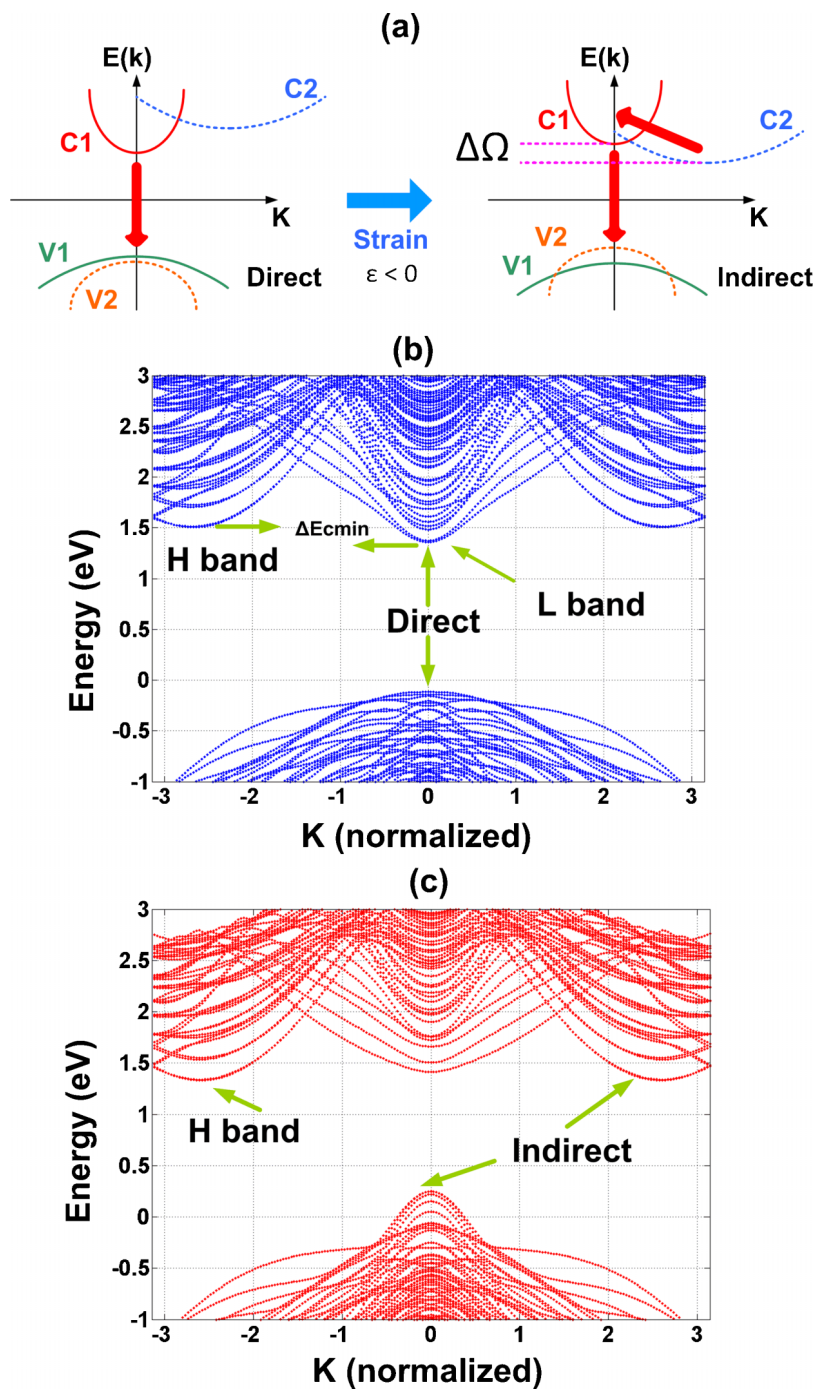

**Figure 3 | Direct to indirect bandgap conversion.** (a) Mechanism of bandgap conversion in response to compressive strain is due to lowering of C2 sub band. Electronic structure of (b) unstrained and (c) −5% strained 3.1 nm [110] SiNW. The indirect conduction band minimum of strained case (c) is lower than the direct conduction band minimum by $\Delta\Omega = 80$ meV. This quantity determines the order of phonon mediated process for light emission.

$$\frac{1}{\tau_{spon}} = \frac{e^2 D^2 n_r}{\epsilon_0 \rho c^3 v_s \hbar} \frac{1}{48\pi^4} \sum \omega_{fm}^2 \left| \langle \psi_f | \hat{r} | \psi_m \rangle \right|^2 \quad (4)$$
$$F(k_f) \cdot Ph_{LA}(k_f, k_i) \Delta k_f$$

D, $\rho$ and $v_s$ represent electron deformation potential (D = 9.5 eV), mass density ($\rho$ = 2329 Kg/m$^3$) and velocity of sound in silicon (9.01 × 10$^5$ cm/sec), respectively[39,40]. $\omega_{fm} = \Delta E_{fm}/\hbar$, where $\Delta E_{fm}$ is energy difference between final and intermediate state. $Ph_{LA}(k_f, k_i)$ contains integration over all possible transverse phonon wave vectors ($q_t$), Bose-Einstein occupancy factors and matrix element of terms like $e^{i\mathbf{q}\cdot\mathbf{r}}$ i.e. $\langle \Psi_m | e^{i\mathbf{q}\cdot\mathbf{r}} | \Psi_i \rangle$ (see **Supplementary Information**). The electron wave vectors at initial (final) states are represented by $k_i$ ($k_f$). $\Delta k_f$ is

$k_z$-space grid element (resolution) along the BZ of the nanowire. This equation is evaluated numerically. The corresponding equation for spontaneous emission time by including LO phonon is given by

$$\frac{1}{\tau_{spon}} = \frac{e^2 |D_{op}|^2 nr}{\epsilon_0 \rho c^3 \hbar \omega_0} \cdot \frac{1}{48\pi^4} \sum \omega_{fm}^2 \left| \langle \psi_f | \hat{r} | \psi_m \rangle \right|^2 \quad (5)$$
$$F(k_f) \frac{E_c(k_i) - E_v(k_f) \pm \hbar \omega_0}{(E_c(k_f) - E_c(k_i) \pm \hbar \omega_0)^2} Ph_{LO}(k_f) \Delta k_f$$

where a dispersion-less optical phonon energy of $\hbar\omega_0 = 63$ meV has been assumed. $Ph_{LO}(k_f)$ contains integration over all possible LO phonon wave vectors ($q_t$), Bose-Einstein occupancy factors and matrix element of terms like e$^{iq.r}$ i.e. $<\Psi_m|e^{iq.r}|\Psi_i>$ (see **Supplementary Information**). $D_{op}$ is the electron deformation potential for LO phonon ($D_{op} = 11 \times 10^8$ eV/cm)[39,40]. $E_c$ and $E_v$ represent the energy of conduction and valence states at a given wave vector ($k_i$ or $k_f$), respectively.

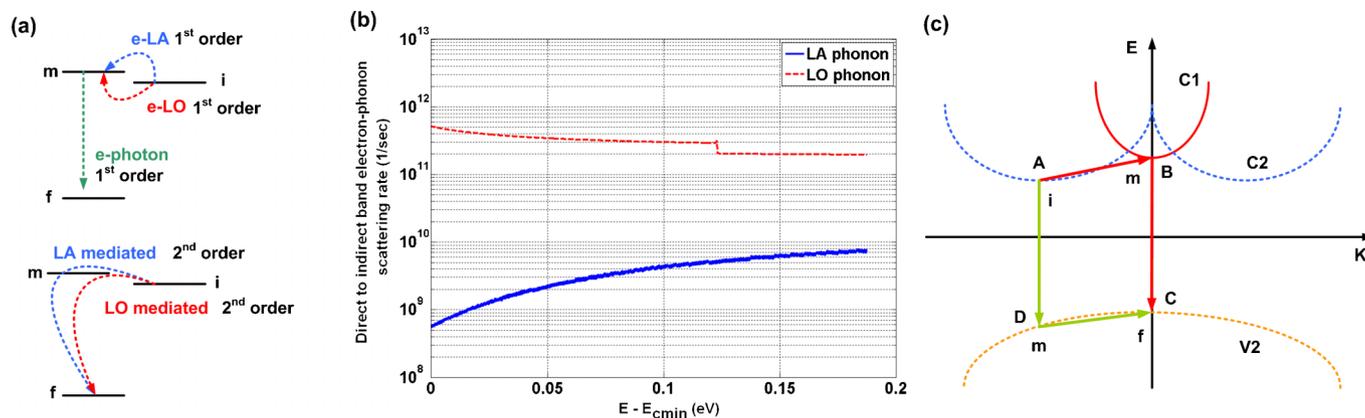

**Figure 4 | First and second order phonon mediated transition.** (a) Top: Two serial 1$^{st}$ order processes which model the light emission calculation in indirect bandgap nanowires with $\Delta\Omega < E_{LO/LA}$. Bottom: A model based on a 2$^{nd}$ order process presented for calculation of light emission in indirect nanowires with $\Delta\Omega > E_{LO/LA}$. (b) Inter-sub band (direct to indirect) electron-phonon scattering rate (s$^{-1}$) versus the energy starting from the bottom of direct conduction band ($E - E_{cmin}$). As can be seen scattering times for electron-LO phonon and electron-LA phonon scattering events are in the order of 1ps and 100ps, respectively. (c) Possible second order transitions from indirect conduction sub band to the valence band maximum. As it is discussed in the **Results** section the A→D→C event is less probable than A→B→C. Initial, intermediate and final states are represented by $i$, $m$ and $f$, respectively.

Equations (4) and (5) can further be reduced to semi-analytic equations by making a few simplifying assumptions. First the momentum matrix element between conduction and valence bands can be assumed to be constant around the BZ center (Fig. 1c). Even if this value varies around BZ center, a corresponding drop of off center values of Fermi-Dirac factor, $F(E_f)$, makes this approximation plausible. More simplifications are also possible if the energy of phonon (e.g. $\hbar\omega_0$ in equation (5)) is neglected in comparison with $E_c(k_i) - E_v(k_f) = E_{gindir}$ and $E_c(k_f) - E_c(k_i) = \Delta\Omega$. As a result, equations (4) and (5) can be reduced to a semi-analytic equation as follows

$$\frac{1}{\tau_{spon}} = \beta_{LO/LA} E_{gdir}^2 R_{cv}^2 E_{gindir} \int_{k_f = -\pi/U_c}^{k_f = +\pi/U_c} F(k_f).S_{q_t}^{LO/LA}(k_f).dk_f \quad (6)$$

where the constant value, $R_{cv}$, is the position matrix element between conduction and valence state at BZ center. $E_{gdir}$ and $E_{gindir}$ represent the direct and indirect bandgap values, respectively. $\beta_{LO/LA}$ is a proportionality factor that contains all relevant constants as given in equations (4) and (5). The term $S^{LA/LO}$ represents Ph$_{LA}$ in equation (4) or Ph$_{LO}$ of equation (5) which is multiplied by simplified energy quotient [see equation (S20) of **Supplementary Information**]. The results of the semi-analytic method are less than 30% off from the numerical method. At room temperature (T = 300 K) and $E_{Fermi} = E_{vmax}$, the average spontaneous emission time due to LA phonons is $\tau_{avg} = 3.4 \times 10^3$ s. Including LO phonons results in $\tau_{avg} = 16$ s, which reveals the strong role of LO phonons in the second order light emission process as well. It is noteworthy that increasing the Fermi level or decreasing the temperature will lead to less available empty states in the valence band; hence larger spontaneous emission time than above is expected. Table 2 lists the spontaneous emission time for an indirect bandgap 3.1 nm [110] SiNW at −5% strain for both LA and LO mediated cases. The stronger LO mediated scattering governs the indirect light emission process according to Fig. 4a (bottom panel). Furthermore, the strong optical anisotropy along z direction is evident in Table 2.

## Discussions
Based on the aforementioned results we anticipate that preparing an inverted population of electrons in the indirect sub band and the subsequent application of strain may cause efficient light emission due to the indirect to direct bandgap conversion. To ensure that there are sufficient carriers in the indirect sub band under realistic conditions of multiple sequential electron-phonon scattering events and an applied electric field, we have calculated the relative occupancy of direct and indirect sub bands using Ensemble Monte Carlo (EMC) simulations[41] (for details refer to **Methods** section). Fig. 5a includes the contributions from the first two sub bands which are less than 4 meV apart (C1 and C2 are taken together; see the Inset of Fig. 5a). As can be seen, the occupancy shows only a relatively small variation across the electric field range considered. The occupancy of the indirect sub bands decreases from approximately 95% at low electric fields, to approximately 92% at 25 kV/cm. The factor of ~10 difference between occupancies of indirect and direct sub bands (Fig. 5a) suggests that periodically straining a biased SiNW can induce population inversion and lasing corresponding to indirect to direct bandgap conversion cycle. The observed increase in the occupancy of electrons in the direct sub band at higher electric fields is attributed to the transition of electrons away from the negative $k_z$ sub bands within BZ as they respond to the electric field. This can be observed in Fig. 5b, which depicts the time evolution of the electron distribution function under different electric fields. The lack of a significant change in the distribution with electric field is primarily attributed to the high electron-phonon scattering rate due to LO phonon emission, and is a feature consistent with other small diameter nanomaterials such as single-wall carbon nanotubes (CNT)[42], and unstrained SiNWs[43]. This can be understood by comparing Fig. 5b with Fig. 5c that depicts the total electron-phonon (LA and LO) scattering rates for conduction sub bands C1 and C2. Under an applied electric field, electrons initially near the bottom of the sub bands gain crystal momentum until they reach the peaks associated with the onset are the dominant players. As soon as an electron gains 63 meV of energy, it emits a phonon and makes a transition to either the same indirect sub band bottom with a very high scattering rate, or the opposite indirect sub band bottom. Because of this very high scattering rate, electrons do not gain enough energy to make a transition to the direct sub band bottom. Also, the LO absorption is very low to begin with since it has dropped by 3 orders of magnitude. At 300 K, the direct to indirect transition or vice versa are almost symmetric since the emission (direct to indirect sub band) and absorption (indirect to direct sub band) have almost the same transition time ($10^{-11}$ s). On the other hand at 77 K the absorption rates drop by 3 orders of magnitude (1/rate=$10^{-8}$ s) while emission rates have still the same order as 300 K case i.e. 1/rate=$10^{-11}$ s. This asymmetry means that carriers have more chance to escape from direct sub band and reach to indirect rather than the reverse.

Table 2 | Spontaneous emission time (s) in a 3.1 nm [110] SiNW (at −5% strain with indirect bandgap). * Role of LO phonons is more significant than LA phonons (100 times more) in determining the emission times

| Spontaneous emission time (s) for indirect bandgap nanowire | | |
|---|---|---|
| Polarization | LA phonons included | LO phonons* included |
| $\tau_x$ | $7.5 \times 10^{+5}$ s | $3.5 \times 10^{+3}$ s |
| $\tau_y$ | $5.4 \times 10^{+5}$ s | $1.9 \times 10^{+3}$ s |
| $\tau_z$ | $3.4 \times 10^{+3}$ s | 16.1 s |
| $\tau_{avg}$ | $3.4 \times 10^{+3}$ s | 16.1 s |

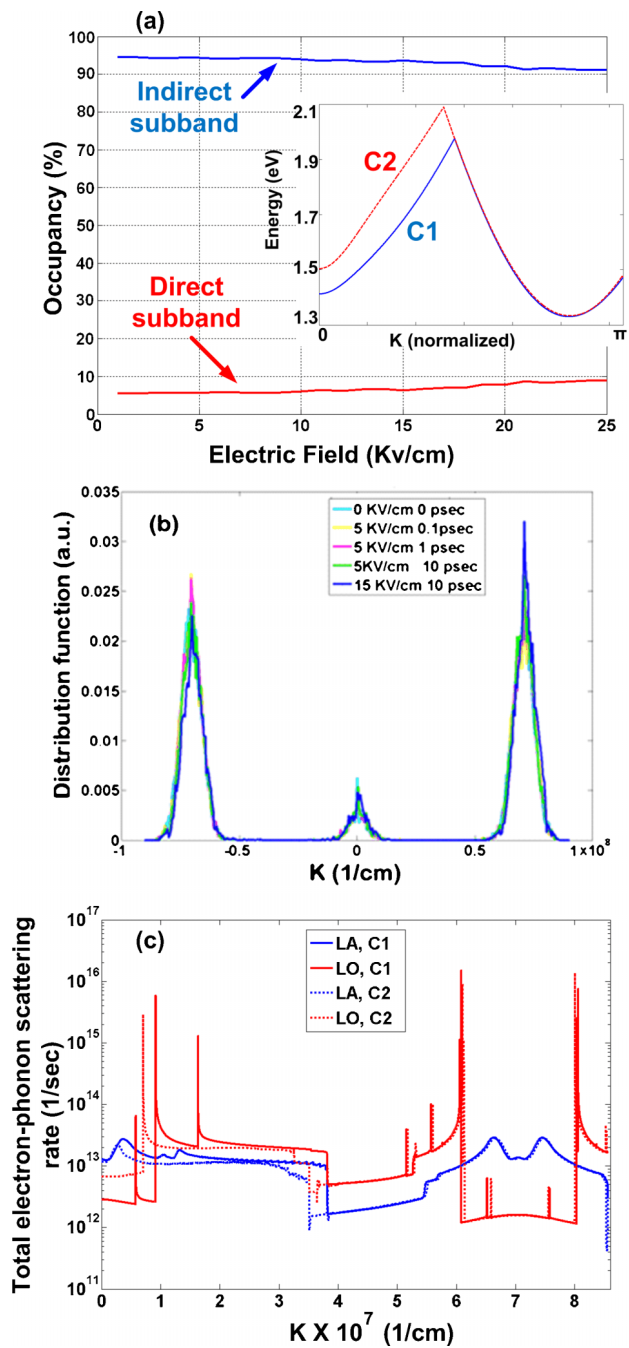

Another non-radiative effect which matters in studying the carrier population is the Auger recombination which can deplete the indirect sub bands. However photoluminescence study of SiNWs with a diameter of 3.3 nm±1.6 nm, as a function of temperature and optical pump intensity has shown[44] that the fastest Auger recombination time can be 19 μs. Since this time is much more than the electron-phonon scattering times that we deal with, we have neglected the Auger recombination effect in our work.

Finally we speculate on how the idea of strain-modulated spontaneous emission time can be developed for SiNWs. There are many examples of recently implemented platforms and methods to apply strain to a single CNT. Piezoelectrically driven table[45], pressure driven deformable substrate[46] and applying force by an atomic force microscope probe[47] are among the methods of generating mechanical strain in CNTs. Additionally there are examples of using a deformable substrate to modulate the energy band diagram of piezoelectric zinc oxide nanowires via strain[48]. Embedding SiNWs on plastic[6], elastomeric[30] and metallic deformable substrates[8] all show that our results are verifiable using existing methods. It is also advantageous that the direct to indirect conversion and the resulting modulation of spontaneous emission time is a reversible process. However the strain value which results from embedding SiNWs in silicon dioxide or nitride shell is unchangeable. The observed population difference in Fig. 5a leads us propose an experiment to observe population inversion and lasing in SiNWs (Fig. 6). If we assume that an indirect bandgap SiNW (under compressive strain) is biased under a moderate electric field, there will be a larger population of electrons in indirect sub band (Fig. 6a). After releasing the strain (or applying tensile strain) to make the nanowire direct bandgap (Fig. 6b), the population which is still significant can scatter to the direct sub band (within picosecond time scales) and stimulated emission takes place, provided that suitable feedback and gain mechanisms have already been designed for such SiNW. It is noteworthy that although strain may cause changes in optical properties of silicon quantum dots, electrical injection and reversible application of strain are practical issues which render the usability of silicon quantum dots difficult.

In summary, we found that *the spontaneous emission time can be modulated by more than one to two orders of magnitude due to strain via two distinct physical mechanisms, a feature that is not observable in bulk silicon crystal. Based on this, we propose an experiment to observe population inversion and lasing in SiNWs.* To more accurately take into account the excitation of carriers from the indirect to direct conduction bands via multiple phonon scattering, we have simulated the nanowire via Ensemble Monte Carlo simulations at room temperature. We found that non-radiative scattering events deplete the initial population of carriers from the indirect sub band. However under moderate electric fields there is still a factor of 10 difference between indirect and direct sub band occupancies. This is of significance when we consider a nanowire device in which lasing and population inversion can be inhibited or initiated by periodic of LO phonon emission. This inelastic-scattering event prevents electrons from gaining further momentum, and induces the observed behaviour in Fig. 5b. To investigate the role of temperature in the aforementioned carrier population analysis we have also conducted simulations for occupancy of the sub bands at 77 K. We observed that the occupancy of the direct part of the sub bands (near BZ center) at 77 K is negligible compared to 300 K. LO phonon emission for inter and intra-sub band are indeed very strong, and application of axial strain. A similar scheme can also be proposed to modulate the photo-absorption of narrow SiNWs, either due to wave function symmetry change or direct to indirect band gap conversion. Change of wave function symmetry can potentially induce large nonlinear optical effects in nanowires in response to strain[49,50] which merits a deeper study. While the calculations are performed for silicon nanowires, similar phenomena should exist in nanowires made out of other material systems.

**Figure 5** | Monte Carlo evolution of sub bands occupancy and electron-phonon scattering rates. (a) Occupancy of indirect and direct sub bands vs. electric field for a 3.1 nm [110] SiNW at −5% strain (indirect bandgap). Inset shows the positive half of the BZ (i.e. k spans [0,π]) with two conduction sub bands used in the EMC simulation. (b) Time evolution of the electron distribution function under different electric fields. (c) The total LA and LO electron-phonon scattering rates. In the legend, C1 and C2 show the initial sub band from which the electron is scattered by emission/absorption of LA and LO phonons.

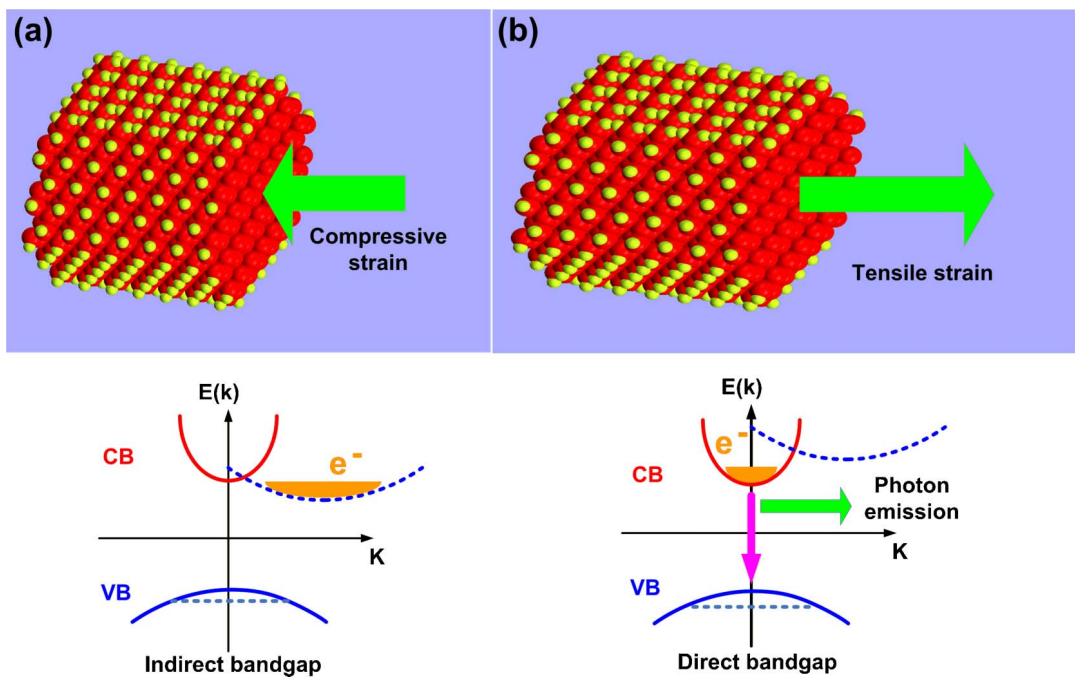

**Figure 6 | A proposed experiment to implement a population inversion in SiNWs.** (a) Injecting current in a compressively strained nanowire (top) which has an indirect bandgap (bottom) can generate an initial population in SiNW. Although non-radiative processes may deplete this sub band of most of those injected carriers, there will still be a factor of 10 difference in the electron occupancy between the indirect and direct sub bands. (b) During strain release or when applying tensile strain (top), the initial population can scatter into the direct sub band via electron-phonon scattering processes within picosecond time scale. Thereafter (bottom) the inverted population can initiate lasing (coherent stimulated emission) if the nanowire is already embedded in a suitable mode enhancing cavity. In the case of incoherent (spontaneous) emission the light has a broader spectrum suitable for light emitting diode (LED) applications. Similar set up can be devised for light absorption (i.e. photocurrent) modulation using strain.

## Methods

The nanowires in this work are cut from bulk silicon crystal in [110] direction and dangling silicon bonds on the surface are terminated with hydrogen atoms (Fig. 1a). The diameters of the nanowires range from 1.7 nm to 3.1 nm. The cross section of nanowires lies in the x-y plane, and z is the axial direction of each nanowire. The relative stability of [110] direction compared to [100], which is quantified as free energy of formation[51], is experimentally verified[33,52]. Energy minimization of nanowires is performed by Density Functional Theory (DFT) method within SIESTA[53] (version 3.1) using Local Density Approximation (LDA) functional with Perdew-Wang (PW91) exchange correlation potential[54]. Spin polarized Kohn-Sham orbitals of double-zeta type with polarization (DZP) have been used. The Brillion Zone (BZ) has been sampled by a set of $1\times1\times40$ k points along the axis of the nanowire (z axis). The minimum center to center distance of SiNWs is assumed to be at least 2 nm to avoid any interaction between nanowires. Energy cut-off, split norm, maximum force tolerance and maximum stress tolerance are 680 eV (50 Ry), 0.15, and 0.01 eV/Å and 1 GPa, respectively. The relaxation stops if the maximum absolute value of interatomic forces is smaller than force tolerance and the maximum stress component is smaller than the stress tolerance. The energy of the unstrained unit cell of the nanowire is minimized using Conjugate Gradient (CG) algorithm during which *variable* unit cell option is selected. Afterwards, for each percent of strain (ε) the unit cell is relaxed using the constant volume chosen by *fixed* unit cell option. With this option atoms are only free to move within the fixed unit cell volume. The result of each minimization step is fed to the next step of minimization. The unit cell length ($U_c$) as defined in Fig. 1a (right) is updated according to the applied strain value ε, i.e. $U_{c-new} = U_{c-old}(1+\varepsilon)$. The electronic structure is calculated within 10 orbital semi-empirical $sp^3d^5s^*$ tight binding scheme[55]. This is to avoid bandgap underestimation due to DFT and diameter sensitive many–body GW corrections. The tight binding method has been successful in regenerating the bulk band structure as well as correctly simulating the boundary conditions i.e. surface passivation[56]. Trend of tight binding bandgap change with diameter in the case of Si nanowire closely matches with the STS experiments[33].

The orbitals of silicon atoms (i.e. s, p, d and s*) are assumed to be of Slater[57] type in which the radial part of each orbital is given by the following equation

$$R(r) = N\ r^{n*-1}\ e^{-\frac{Z-S}{n*}r}$$

where N is the normalization factor and Z is the atomic number. The shielding constant (s) and effective quantum number (n*) are found using the rules given by Slater[57].

To calculate the spontaneous emission life times in direct bandgap nanowires [equation (2) in the main section], Fermi's golden rule with first order perturbation theory[58] is used. The matrix element of electron-photon interaction Hamiltonian ($H_{e-R}$) can be simplified to momentum matrix element, $<\Psi_i|\hat{e}.P|\Psi_f>$. This is further reduced to its position representation and integrals of type $<\alpha(r-R_o)|r|\beta(r-R_o)>$. Here $r$ is the position operator, α and β are atomic orbitals of which $\Psi_i$ and $\Psi_f$ are composed. These integrals have two parts i.e. $R_o\delta_{\alpha\beta} + <\alpha(u)|u|\beta(u)>$ where $R_o$ is the position of the atom. The second part consists of radial and angular integration of Slater type orbitals which are both found analytically using Wolfram Mathematica® online integrator (http://integrals.wolfram.com/index.jsp). Among 100 combinations of 10 orbitals, 15 of them have symmetry-allowed nonzero value. The first order electron-phonon scattering rates are also calculated using Fermi's golden rule within the first order perturbation scheme[58]. The electron-phonon interaction Hamiltonian ($H_{eP}$) is of deformation potential type for bulk LA and LO phonons. As it is stated by M. Nawaz et al[59], taking confined LO phonons into account will reduce the scattering rate. Thus phonon confinement does not have an adverse effect on the spontaneous emission times calculated. Also it is shown that there is only a 10% difference in calculated mobility between the cases where bulk and confined phonons are used[60]. Details of calculating the electron-phonon interaction Hamiltonian matrix elements and scattering rate have been skipped and the interested reader can refer to[39,40]. For indirect bandgap nanowires with larger energy offset (i.e. $\Delta\Omega > E_{LO}/E_{LA}$), second order perturbation method is used in which all interaction Hamiltonian matrix elements are calculated likewise. The expressions for spontaneous emission time are explained in the main section.

To further investigate the carrier population statistics of indirect sub bands under the influence of electric field and multi electron-phonon scattering events, we use standard Ensemble Monte Carlo (EMC) methodology[41]. In setting up our EMC simulation, we consider an infinitely long, defect free SiNW with a uniform temperature. The electric field is also uniform and directed along the axis of the SiNW. In performing the simulation, tabulated values of two lowest conduction sub bands of 3.1 nm diameter [110] SiNW with indirect bandgap (due to −5% strain) (C1 and C2 in the Inset of Fig. 5a) are used. For each initial state starting from indirect conduction sub band minimum, all possible final states within C1 and C2 with corresponding scattering rates for both phonon types (LA/LO) were utilized. Both inter- and intra-sub band electron-phonon scattering events have been calculated. The rest of the conduction sub bands are not included in the simulation since the third conduction sub band is at least 100 meV above the first two sub bands (C1 and C2 in Fig. 5a). Electron transport is confined to the first BZ, which is divided into 8001 k grid points (4000 positive k and 4000 negative k values) and for which the tabulated energy values and electron-phonon scattering rates are computed and stored. Electrons are initially injected into the SiNW at the bottom of an indirect sub band and the simulation is initially executed for 500,000 scattering cycles at 0 kV/cm electric field so as to allow the electrons to approach as close to an equilibrium distribution as possible.

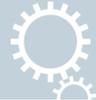

### Acknowledgements

Daryoush Shiri acknowledges the 2008, 2009 and 2010 Nanofellowship awards from The Waterloo Institute for Nanotechnology (WIN). Access to the facilities of the Shared Hierarchical Academic Research Computing Network (SHARCNET) of Ontario, Canada is also acknowledged. Amit Verma acknowledges the use of the High Performance Computing resource at the Texas A&M University-Kingsville supported through NSF Grant No. NSF-0619810, for Monte Carlo simulations. We gratefully acknowledge support from National Science Foundation under Grant No. 1001174.

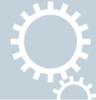

## Author contributions

D. S performed DFT and tight binding calculations of electronic structure of nanowires, mathematical derivations of spontaneous emission times, electron-phonon scattering rates and coding. A.V wrote the ensemble Monte Carlo code and calculated the evolution of carrier populations. C. R. S was involved in the study of the problem and device design point of view of simulation data. M. P. A defined the research problem and idea of strain application to nanowire. Manuscript was written by D. S, A.V and M. P. A and they contributed equally in discussions and manuscript review.

## Additional information

**Supplementary information** accompanies this paper at http://www.nature.com/scientificreports

**Competing financial interests:** The authors declare no competing financial interests.

**License:** This work is licensed under a Creative Commons Attribution-NonCommercial-NoDerivative Works 3.0 Unported License. To view a copy of this license, visit http://creativecommons.org/licenses/by-nc-nd/3.0/

**How to cite this article:** Shiri, D., Verma, A., Selvakumar, C.R. & Anantram, M.P. Reversible Modulation of Spontaneous Emission by Strain in Silicon Nanowires. *Sci. Rep.* **2**, 461; DOI:10.1038/srep00461 (2012).